\documentclass[final,1p,times]{elsarticle} 

\usepackage{graphicx}
\usepackage{amssymb} 
\usepackage{url}

\journal{Nuclear Physics A} 

\begin{document}

\begin{frontmatter} 

\title{EPS09s and EKS98s: Impact parameter dependent nPDF sets}

\author[jyu,hip]{I.~Helenius}
\author[jyu,hip]{K.J.~Eskola}
\author[psu]{H.~Honkanen}
\author[sdg,cern]{C.A.~Salgado}

\address[jyu]{Department of Physics, P.O. Box 35, FI-40014 University of Jyv\"askyl\"a, Finland}
\address[hip]{Helsinki Institute of Physics, P.O. Box 64, FIN-00014 University of Helsinki, Finland}
\address[psu]{The Pennsylvania State University, 104 Davey Lab, University Park, PA 16802, USA}
\address[sdg]{Departamento de F\'\i sica de Part\'\i culas and IGFAE, Universidade de Santiago de Compostela, Galicia-Spain}
\address[cern]{Physics Department, Theory Unit, CERN, CH-1211 Gen\`eve 23, Switzerland}

\begin{abstract} 
In our recent study we have determined two new spatially dependent nuclear PDF (nPDF) sets, EPS09s and EKS98s. With these, the hard-process cross-sections can be calculated in different centrality classes consistently with the globally analyzed nPDFs for the first time.
The sets were determined by exploiting the $A$-systematics of the globally fitted nPDF sets, EPS09 and EKS98. For the spatial dependence of the nPDFs we used a power series ansatz in the nuclear thickness function $T_A$. In this flash talk we introduce the framework, and present our NLO EPS09s-based predictions for the nuclear modification factor in four centrality classes for inclusive neutral pion production in p+Pb collisions at the LHC
and for inclusive prompt photon production in d+Au collisions at RHIC at midrapidity.
\end{abstract} 

\end{frontmatter} 


\section{Introduction}

According to the collinear factorization theorem \cite{Collins:1989gx,Brock:1993sz} 
the inclusive yield of a parton $k$ in a heavy ion collision at a given impact parameter $\mathbf{b}$ can be calculated from
\begin{equation}
\mathrm{d} N^{AB\rightarrow k+X} = T_{AB}(\mathbf{b})\sum_{i,j,X'}f_i^A(x_1,Q^2)\otimes f_j^B(x_2,Q^2) \otimes \mathrm{d}\hat{\sigma}^{ij\rightarrow k+X'} + \mathcal{O}(1/Q^2),
\end{equation}
where $T_{AB}(\mathbf{b})$ is the nuclear overlap function, $f_i^A(x_1,Q^2)$ ($f_j^B(x_2,Q^2)$) is a universal, process-independent nuclear parton distribution function (nPDF) describing the momentum distribution of a parton $i$ ($j$) inside the nucleus $A$ ($B$), and $\mathrm{d}\hat{\sigma}^{ij\rightarrow k+X'}$ can be computed using perturbative QCD. The nPDFs are known to be modified with respect to corresponding free proton PDFs and this modification can be quantified by defining the ratio $R_i^A(x,Q^2)$ as
\begin{equation}
R_i^A(x,Q^2) \equiv f_i^A(x,Q^2)/f_i^p(x,Q^2).
\end{equation}
As the (n)PDFs are of non-perturbative origin, the nuclear modifications of the PDFs cannot be computed from the first principles of QCD, but these can be determined by global fitting to the experimental data using DGLAP evolution as done e.g. in Refs.~\cite{Eskola:2009uj, Eskola:1998df}. So far all the global fits have considered only minimum bias observables and thus the resulting nPDFs should be considered as spatially averaged quantities. This implies that one has not been able to consistently compute the hard-process cross sections in different centrality classes. In \cite{Helenius:2012wd} we have considered this problem and developed a framework for spatial dependence of the nPDFs.

\section{Framework}

\begin{figure}[htb]
\begin{minipage}[t]{0.49\linewidth}
\centering
\includegraphics[width=\textwidth]{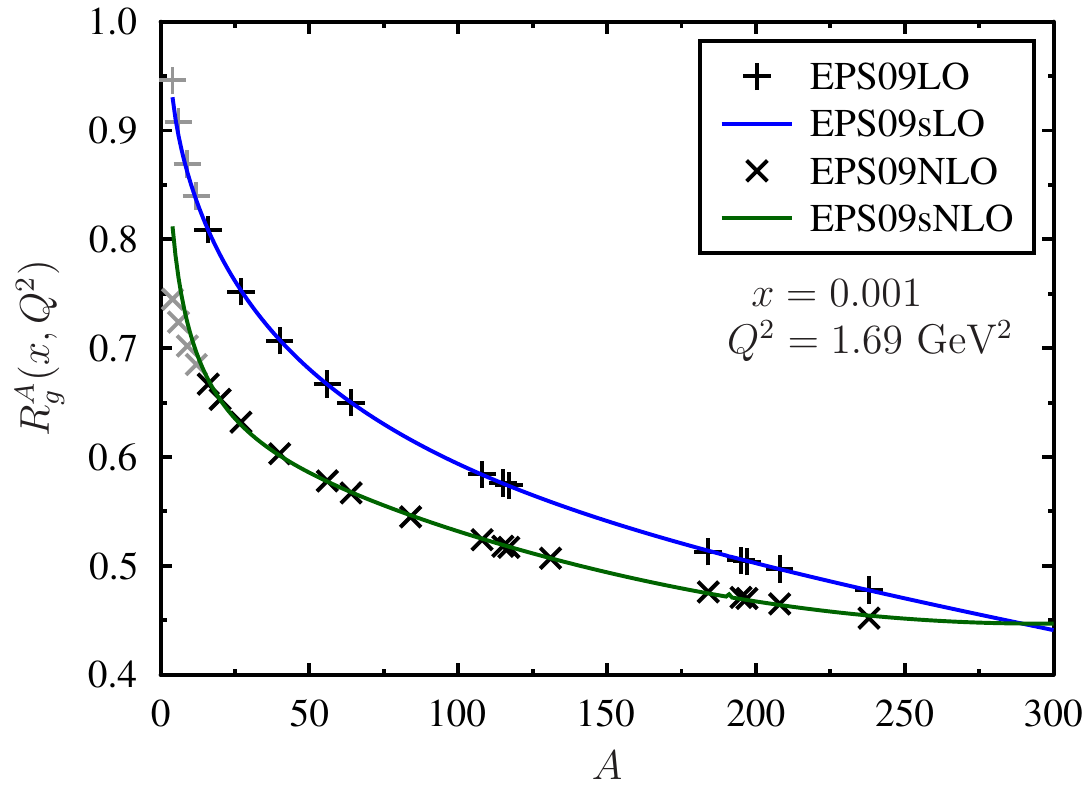}
\vspace{-15pt}
\caption{The spatially averaged gluon modification $R_g(x,Q^2)$ from EPS09 and EPS09s in LO and NLO as a function of $A$ for fixed values of $x$ and $Q^2$. From \cite{Helenius:2012wd}.}
\label{fig:R_g_A_fits}
\end{minipage}
\hspace{0.02\linewidth}
\begin{minipage}[t]{0.49\linewidth}
\centering
\includegraphics[trim = 10pt 0pt 0pt 0pt, clip, width=\textwidth]{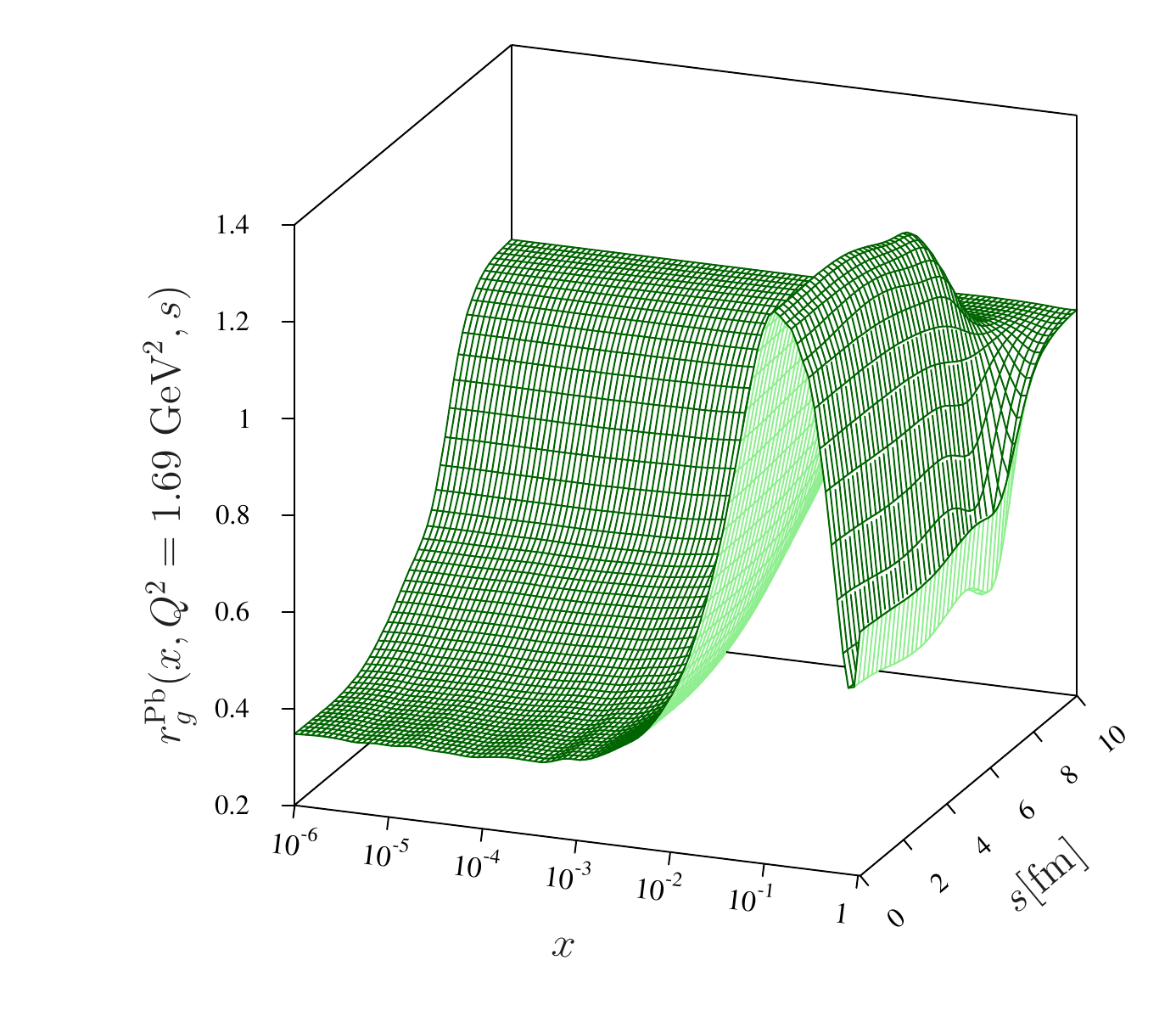}
\vspace{-15pt}
\caption{The spatially dependent gluon modification $r_g(x,Q^2,\mathbf{s})$ from EPS09sNLO for $A=208$ as a function of $x$ and $s$ at the initial scale of EPS09. From \cite{Helenius:2012wd}.}
\label{fig:R_g_3d}
\end{minipage}
\vspace{-8pt}
\end{figure}

We introduce a nuclear modification $r_i^A(x,Q^2,\mathbf{s})$ which now depends also on the transverse position $\mathbf{s}$ in the nucleus. For this we require that the spatial average is identical to the globally analyzed nuclear modification
\vspace{-9pt}
\begin{equation}
R_{i}^{A}(x,Q^2) \equiv \frac{1}{A}\int \mathrm{d}^2 \mathbf{s} \,T_A(\mathbf{s}) \, r_{i}^{A}(x,Q^2,\mathbf{s}),
\vspace{-3pt}
\end{equation}
which we take here from the earlier global analysis, EPS09 \cite{Eskola:2009uj} or EKS98 \cite{Eskola:1998df}. As this constrains only the spatially averaged modification, we need to make an assumption about the spatial dependence. Motivated by the small-$x$ shadowing studies \cite{Frankfurt:2011cs}, we assume that the nuclear modification is related to the nuclear thickness function $T_A(\mathbf{s})$ and choose a power series form
\vspace{-3pt}
\begin{equation}
r_i^A(x,Q^2,\mathbf{s}) = 1 + \sum_{j=1}^{n} c_{j}^{i}(x,Q^2)\,[T_A(\mathbf{s})]^j,
\label{eq:series}
\vspace{-4pt}
\end{equation}
where $c_{j}^{i}(x,Q^2)$'s are now $A$-independent fit parameters. 
We determine the values of these parameters by requiring the same $A$-dependence for the spatially averaged modification as obtained in the earlier global analysis, i.e. we minimize the $\chi^2$ defined as
\vspace{-3pt}
\begin{equation}
\chi^2_i(x,Q^2) = \sum_A \left[\frac{R^{A}_{i}(x,Q^2) - \frac{1}{A}\int \mathrm{d}^2 \mathbf{s}\, T_A(\mathbf{s})\,r^{A}_{i}(x,Q^2,\mathbf{s})}{W^{A}_{i}(x,Q^2)} \right]^2.
\end{equation}
For the weight factor we use $W^{A}_{i}(x,Q^2) = 1$ for EPS09 and $W^{A}_{i}(x,Q^2) = 1-R^{A}_{i}(x,Q^2)$ for EKS98. In figure~\ref{fig:R_g_A_fits} we plot the spatially averaged gluon modification as a function of $A$ from EPS09LO and EPS09NLO and from the corresponding spatial fits for fixed $x$ and $Q^2$ values. We observe that taking $n=4$ in equation~(\ref{eq:series}) is sufficient to reproduce the $A$-systematics accurately for all the sets considered. By repeating this fitting procedure over all parton flavors and over an $(x,Q^2)$ grid for all different sets (EKS98: LO, EPS09: LO and NLO + 30 error sets each) we obtain the new spatially dependent nPDF sets \texttt{EPS09s} and \texttt{EKS98s}, which are now available at our webpage\footnote{\url{https://www.jyu.fi/fysiikka/en/research/highenergy/urhic/nPDFs}}. In figure~\ref{fig:R_g_3d} we plot the gluon modification from EPS09sNLO as a function of $x$ and $s\,(=|\mathbf{s}|)$ at the EPS09 initial scale $Q^2=1.69\,\rm{GeV}^2$. 
From these two figures one can notice how the $A$-dependence of the $R_i^A(x,Q^2)$ maps to the spatial depedence of the $r_i^A(x,Q^2,\mathbf{s})$: modifications are larger at the thick center of nucleus and decrease towards the thin edge.


\section{Applications}

\begin{figure}[b!]
\vspace{-12pt}
\centering
\includegraphics[width=0.95\textwidth]{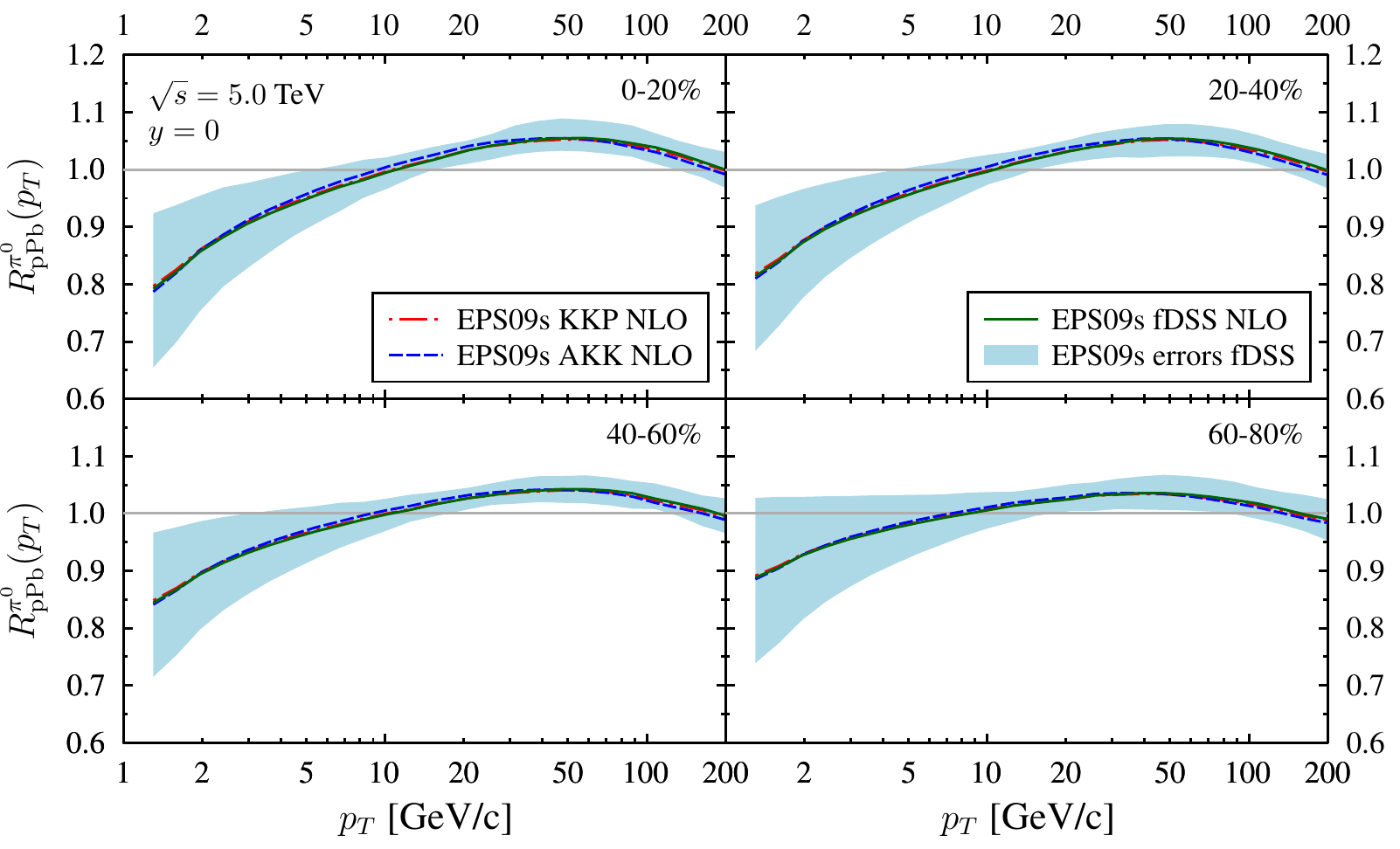}
\vspace{-5pt}
\caption{The nuclear modification factor $R_{\rm pPb}^{\pi^0}$ for inclusive neutral pion production in p+Pb collisions at the LHC for four different centrality classes (different panels) at midrapidity, calculated with three different FFs, KKP (dot-dashed), AKK (dashed) and fDSS (solid). The blue uncertainty band is calculated from the 30 error sets in EPS09s with fDSS FFs. From \cite{Helenius:2012wd}.}
\label{fig:R_pPb_pi0}
\vspace{-5pt}
\end{figure}

In Ref.~\cite{Helenius:2012wd} we compared our NLO calculation of the nuclear modification factor for $\pi^0$ production in d+Au collisions at $y=0$ and $\sqrt{s_{NN}}=200\,\mathrm{GeV}$ with the PHENIX measurement \cite{Adler:2006wg} in four centrality classes. Within the uncertainties, the calculations described the data well in all these centrality classes. 
In figure~\ref{fig:R_pPb_pi0} we plot similar calculation for p+Pb collisions at $\sqrt{s_{NN}} = 5.0\,\mathrm{TeV}$ and $y=0$ in nucleon-nucleon CMS-frame at the LHC. The centrality classes we consider here are $0-20\,\%$, $20-40\,\%$, $40-60\,\%$, and $60-80\,\%$. These are determined in terms of impact parameter intervals which we calculate using the optical Glauber model, see Ref.~\cite{Helenius:2012wd} for details. The calculations are done with the \texttt{INCNLO}-package\footnote{\url{http://lapth.in2p3.fr/PHOX_FAMILY/readme_inc.html}} \cite{Aversa:1988vb} and CTEQ6M \cite{Pumplin:2002vw} free proton PDFs with the EPS09s modifications, and with three fragmentation functions (FFs): KKP \cite{Kniehl:2000fe}, AKK \cite{Albino:2008fy}, and fDSS \cite{deFlorian:2007aj}. 
Our result for the corresponding minimum-bias ratio $R_{\rm pPb}^{\pi^0}(p_T)$ (left panel of figure~17 in Ref.~\cite{Helenius:2012wd}) agrees encouragingly well with the first ALICE data for $R_{\rm pPb}^{\rm ch}(p_T)$ \cite{ALICE:2012mj}. From figure~\ref{fig:R_pPb_pi0} we observe that the nuclear effects are more pronounced in the most central collisions and they decrease towards more peripheral collisions. Also, we see that the possible differences between different FFs essentially cancel out in ratios like $R_{\rm pPb}^{\pi^0}(p_T)$.

\begin{figure}[htb!]
\vspace{-15pt}
\centering
\includegraphics[width=0.95\textwidth]{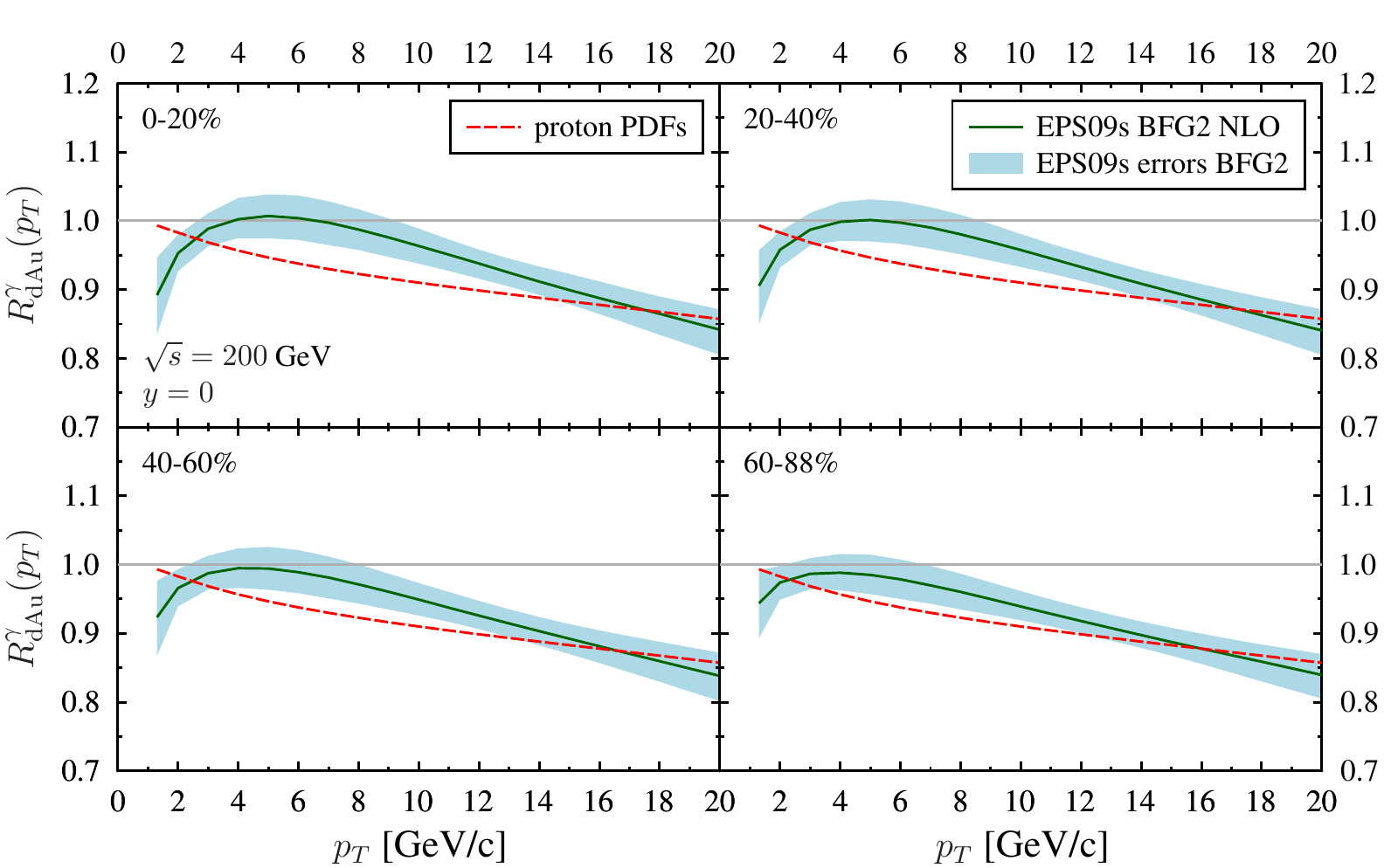}
\vspace{-9pt}
\caption{The nuclear modification factor $R_{\rm dAu}^{\gamma}$ for inclusive prompt photon production in d+Au collisions at RHIC for four centrality classes (different panels) at midrapidity with and without the nuclear modification of the PDFs. Fragmentation component is calculated with BFG FFs and the blue uncertainty band is calculated from the 30 error sets in EPS09s. From \cite{HeleniusEskola}.}
\vspace{-15pt}
\label{fig:R_dAu_gamma}
\end{figure}

In figure \ref{fig:R_dAu_gamma} we plot the nuclear modification factor for inclusive prompt (= direct + fragmentation) photon production in d+Au collisions at $\sqrt{s_{NN}}=200\,\mathrm{GeV}$ and $y=0$. For the parton fragmentation to photons we have used the BFG (set II) FFs \cite{Bourhis:1997yu}. To quantify the isospin effect we have plotted also the ratio $R_{\rm dAu}^{\gamma}$ without the nuclear modifications to each panel. At $p_T\sim 5\,\mathrm{GeV/c}$ the antishadowing in the nPDFs seems to compensate for the isospin suppression and at $p_T\sim 15-20\,\mathrm{GeV/c}$ the isospin suppression dominates so the centrality dependence of $R_{\rm dAu}^{\gamma}$ is minimal at this region.

\section*{Acknowledgements}
\vspace{-4pt}
I.H. and K.J.E. thank the Magnus Ehrnrooth Foundation and the Academy of Finland (Project 133005) for financial support. C.A.S. is supported by the European Research Council grant HotLHC ERC-2011-StG-279579, Ministerio de Ciencia e Innovaci\'on of Spain, and Xunta de Galicia.
H.H. is supported by the U.S. Department of Energy under Grant DE-FG02-93ER40771.

\vspace{-10pt}
\bibliographystyle{model1a-num-names}
\bibliography{QM2012_proc}

\begin{thebibliography}{15}
\expandafter\ifx\csname natexlab\endcsname\relax\def\natexlab#1{#1}\fi
\providecommand{\bibinfo}[2]{#2}
\ifx\xfnm\relax \def\xfnm[#1]{\unskip,\space#1}\fi
\bibitem[{Collins et~al.(1988)Collins, Soper, and Sterman}]{Collins:1989gx}
\bibinfo{author}{J.~C. Collins}, \bibinfo{author}{D.~E. Soper},
  \bibinfo{author}{G.~F. Sterman}, \bibinfo{journal}{Adv.Ser.Direct.High Energy
  Phys.} \bibinfo{volume}{5} (\bibinfo{year}{1988}) \bibinfo{pages}{1--91}.
\bibitem[{Brock et~al.(1995)}]{Brock:1993sz}
\bibinfo{author}{R.~Brock}, et~al., \bibinfo{journal}{Rev.Mod.Phys.}
  \bibinfo{volume}{67} (\bibinfo{year}{1995}) \bibinfo{pages}{157--248}.
\bibitem[{Eskola et~al.(2009)Eskola, Paukkunen, and Salgado}]{Eskola:2009uj}
\bibinfo{author}{K.~J. Eskola}, \bibinfo{author}{H.~Paukkunen},
  \bibinfo{author}{C.~A. Salgado}, \bibinfo{journal}{JHEP} \bibinfo{volume}{04}
  (\bibinfo{year}{2009}) \bibinfo{pages}{065}.
\bibitem[{Eskola et~al.(1999)Eskola, Kolhinen, and Salgado}]{Eskola:1998df}
\bibinfo{author}{K.~J. Eskola}, \bibinfo{author}{V.~J. Kolhinen},
  \bibinfo{author}{C.~A. Salgado}, \bibinfo{journal}{Eur.Phys.J.}
  \bibinfo{volume}{C9} (\bibinfo{year}{1999}) \bibinfo{pages}{61--68}.
\bibitem[{Helenius et~al.(2012)Helenius, Eskola, Honkanen, and
  Salgado}]{Helenius:2012wd}
\bibinfo{author}{I.~Helenius}, \bibinfo{author}{K.~J. Eskola},
  \bibinfo{author}{H.~Honkanen}, \bibinfo{author}{C.~A. Salgado},
  \bibinfo{journal}{JHEP} \bibinfo{volume}{1207} (\bibinfo{year}{2012})
  \bibinfo{pages}{073}.
\bibitem[{Frankfurt et~al.(2012)Frankfurt, Guzey, and
  Strikman}]{Frankfurt:2011cs}
\bibinfo{author}{L.~Frankfurt}, \bibinfo{author}{V.~Guzey},
  \bibinfo{author}{M.~Strikman}, \bibinfo{journal}{Phys.Rept.}
  \bibinfo{volume}{512} (\bibinfo{year}{2012}) \bibinfo{pages}{255--393}.
\bibitem[{Adler et~al.(2007)}]{Adler:2006wg}
\bibinfo{author}{S.~S. Adler}, et~al., \bibinfo{journal}{Phys.Rev.Lett.}
  \bibinfo{volume}{98} (\bibinfo{year}{2007}) \bibinfo{pages}{172302}.
\bibitem[{Aversa et~al.(1989)Aversa, Chiappetta, Greco, and
  Guillet}]{Aversa:1988vb}
\bibinfo{author}{F.~Aversa}, \bibinfo{author}{P.~Chiappetta},
  \bibinfo{author}{M.~Greco}, \bibinfo{author}{J.~P. Guillet},
  \bibinfo{journal}{Nucl. Phys.} \bibinfo{volume}{B327} (\bibinfo{year}{1989})
  \bibinfo{pages}{105}.
\bibitem[{Pumplin et~al.(2002)Pumplin, Stump, Huston, Lai, Nadolsky
  et~al.}]{Pumplin:2002vw}
\bibinfo{author}{J.~Pumplin}, \bibinfo{author}{D.~R. Stump},
  \bibinfo{author}{J.~Huston}, \bibinfo{author}{H.~L. Lai},
  \bibinfo{author}{P.~M. Nadolsky}, et~al., \bibinfo{journal}{JHEP}
  \bibinfo{volume}{0207} (\bibinfo{year}{2002}) \bibinfo{pages}{012}.
\bibitem[{Kniehl et~al.(2000)Kniehl, Kramer, and Potter}]{Kniehl:2000fe}
\bibinfo{author}{B.~A. Kniehl}, \bibinfo{author}{G.~Kramer},
  \bibinfo{author}{B.~Potter}, \bibinfo{journal}{Nucl.Phys.}
  \bibinfo{volume}{B582} (\bibinfo{year}{2000}) \bibinfo{pages}{514--536}.
\bibitem[{Albino et~al.(2008)Albino, Kniehl, and Kramer}]{Albino:2008fy}
\bibinfo{author}{S.~Albino}, \bibinfo{author}{B.~A. Kniehl},
  \bibinfo{author}{G.~Kramer}, \bibinfo{journal}{Nucl.Phys.}
  \bibinfo{volume}{B803} (\bibinfo{year}{2008}) \bibinfo{pages}{42--104}.
\bibitem[{de~Florian et~al.(2007)de~Florian, Sassot, and
  Stratmann}]{deFlorian:2007aj}
\bibinfo{author}{D.~de~Florian}, \bibinfo{author}{R.~Sassot},
  \bibinfo{author}{M.~Stratmann}, \bibinfo{journal}{Phys.Rev.}
  \bibinfo{volume}{D75} (\bibinfo{year}{2007}) \bibinfo{pages}{114010}.
\bibitem[{Abelev et~al.(2012)}]{ALICE:2012mj}
\bibinfo{author}{B.~Abelev}, et~al., \bibinfo{journal}{arXiv:1210.4520
  [hep-ex]}  (\bibinfo{year}{2012}).
\bibitem[{Bourhis et~al.(1998)Bourhis, Fontannaz, and Guillet}]{Bourhis:1997yu}
\bibinfo{author}{L.~Bourhis}, \bibinfo{author}{M.~Fontannaz},
  \bibinfo{author}{J.~Guillet}, \bibinfo{journal}{Eur.Phys.J.}
  \bibinfo{volume}{C2} (\bibinfo{year}{1998}) \bibinfo{pages}{529--537}.
\bibitem[{Helenius and Eskola(2012)}]{HeleniusEskola}
\bibinfo{author}{I.~Helenius}, \bibinfo{author}{K.~J. Eskola},
  \bibinfo{journal}{work in progress}  (\bibinfo{year}{2012}).

\end{thebibliography}

\end{document}